\documentclass[sigplan,screen]{acmart}
\AtBeginDocument{%
  }

\copyrightyear{2025}
\acmYear{2025}
\setcopyright{cc}
\setcctype{by-nc-sa}
\acmConference{Access InContext Workshop @ CHI’25}{April 26}{Yokohama, Japan}
\acmBooktitle{Access InContext Workshop @ CHI’25, April 26, 2025, Yokohama, Japan}

\begin{document}

\title{The Spectacle of \textit{Fidelity}: Blind Resistance and the Wizardry of Prototyping}


\author{Hrittika Bhowmick}
\affiliation{%
 \institution{BITS-Pilani, Hyderabad Campus}
 \city{Hyderabad}
 \country{India}
 }

\author{Shilpaa Anand}
\affiliation{%
  \institution{BITS-Pilani, Hyderabad Campus}
  \city{Hyderabad}
  \country{India}}

\renewcommand{\shortauthors}{BHOWMICK, H. AND ANAND, S.}

\begin{abstract}
 Prototyping is widely regarded in Human–Computer Interaction (HCI) as an iterative process through which ideas are tested and refined, often via visual mockups, screen flows, and coded simulations. This position paper critiques the visual-centric norms embedded in prototyping culture by drawing from the lived experiences of blind scholars and insights from cultural disability studies. The review discusses how dominant methods of prototyping rely on an unexamined \textit{fidelity} to sight: privileging what can be rendered visibly coherent while marginalizing other modes of knowing and making. By repositioning prototyping as a situated, embodied, and relational practice, this paper challenges HCI to rethink what kinds of design participation are legitimized and which are left out, when prototyping is reduced to screen-based simulations. 
\end{abstract}



\maketitle

\section{Introduction}
“Pay no attention to that man behind the curtain” — \textit{The Wonderful Wizard of Oz}
\\\\In L. Frank Baum’s early 20th-century fairytale \cite{bernsen1998wizard}, \textit{The Wonderful Wizard of Oz}, this command marks the moment when illusion collapses. The supposedly omniscient wizard is revealed not as a magical being but as a nervous balloonist, hidden behind levers, mirrors, and voice projection devices, \textit{performing authority} through spectacle. Dorothy and her companions, the Tin Woodman, the Scarecrow, and the Lion, each marked by a perceived lack (a home, a heart, a brain, and courage), had embarked on a dangerous quest believing that someone with more knowledge and power could fix them. What they find instead is the machinery of make-belief, built on appearances, not care. This moment resonates beyond fantasy. Dramatizes a politics of perception, and not on mutual recognition but on controlled visibility. In Human-Computer Interaction (HCI), such dynamics echo in the Wizard of OZ (WOZ) prototyping method, where designers simulate inbuilt technologies behind the scenes, observing \textit{users} who interact with interfaces believed to be autonomous \cite{bernsen1998wizard}. Like Baum’s Great Wizard of Oz, the designer manipulates not just functionality, but beliefs of the interlocutors.

Prototypes are often designed with the assumption that they must be \textit{seen} to be understood, their fidelity measured in pixels and polish. But this assumption of vision as the default modality produces a design culture where accessibility becomes retroactive, and alternative forms of perception are treated as deviations. This paper takes up the moment of revelation in fairytale as a methodological and epistemic critique. Drawing from the lived experiences of blindness documented within cultural disability studies scholarship, it challenges the ocularnormative foundations of prototyping in Human-Computer Interaction (HCI). In centering blind epistemologies, we ask: \textit{What forms of fidelity emerge when design is approached through non-sighted ways of knowing?} Re-framing accessibility as an ongoing, situated negotiation, we apply this lens to prototyping culture itself \cite{williamson2019accessible}. In doing so, the paper positions Access InContext as a rethinking of design’s very conditions of perception, iteration, and legibility that re-imagines the practice of prototyping altogether.

\section{Blind Epistemologies and the Re-imagination of Prototyping}
The paper’s positional stance on critiquing sightedness is not an attempt to elevate blindness over other disability experiences. It rather responds to a specific methodological context in interaction design, where vision is structurally assumed and rarely interrogated. As outlined in the introduction, sight functions merely as a sense and more as a disciplining logic, shaping how prototyping is taught and practiced, which implicitly divides roles: the \textit{designer} and the \textit{respondent} (widely debated as \textit{user}). This split enacts deeper assumptions about authorship and feedback, where the designer (typically sighted) drafts sketches, wireframes, or design mock-ups, while the respondent (often imagined as blind or disabled) reacts to what is already fixed, visually. Industry-standard tools such as Figma, AdobeXD and Sketch are built for visual grammars: spatial hierarchies, color-coded logic, drag-and-drop interfaces, and screen-based interactions. These are not inherently hostile to non-visual design logics, but they certainly do not center them \cite{li2021accessibility}. Visual prototypes are treated as universal, a lingua franca of design, while alternative forms of knowing and communicating remain illegible. This \textit{myth} reinforces a logic in which feedback is visualized and usability is quantified through observation. With this in mind, the following themes of the literature review take up two critical sites of inquiry — \textit{visualization} and \textit{observation}, to examine how prototyping performs cultural work. 
\subsection{From Literalism to Lived Modalities}
The foundation for understanding blind disability studies scholars’ critique of visual culture begins with recognizing that sight itself is not a neutral biological function but a socially and politically constructed way of knowing. As Hemachandran Karah argues, “seeing is as much a political act as it pertains to human bonding,” \cite{epwStaresBlind} challenging the assumption that visual perception is purely physiological and repositioning it as a social practice embedded in power relations. Sight, in this formulation, becomes both a literal and symbolic structure of epistemological legitimacy. Importantly, blind scholars have also turned their “gaze” back onto sightedness as a particular sensory ideology. Karah and Michalko both suggest that the experience of blindness makes visible the narrow contours of what is taken for granted in sighted visuality \cite{epwStaresBlind,michalko2001blindness}. The emphasis on clarity, speed, and representational precision, values central to visual design and by extension, to prototyping, are experienced by blind individuals as mechanisms of exclusion. 

In this context, Prototyping becomes one of the clearest arenas where sighted epistemologies are performed and institutionalized. The very idea of making something “visible” for feedback presumes a visual audience. Even the notion of iterating on a design often presumes changes that are representable through visual refinements. However, drawing from the critical lens of blind scholarship, prototyping must be re-understood as a way of making ideas perceptible. Blind epistemologies compel us to imagine futures where ideas emerge relationally: through conversation, shared orientation in space, co-constructed rhythms of sound, touch, and movement which transform prototypes into a generative site for social meaning-making. For instance,\citeauthor{michalko2001blindness} recounts a moment in his classroom when he engaged with a sighted student in a conversation that was not predicated on the visual cues of gaze or gesture, but on tone and mutual narrative filled with humor that improved the student-teacher exchange despite the hidden discomfort and amusement within the sighted classroom. The classroom became, in effect, a prototype for Michalko to reflect on later as a relational space of testing and refining pedagogical ideas that neither required nor privileged sight. 

Such instances reveal that blindness enables alternative ways of “sketching” ideas into being, practices that resist the crude literalism of visual mock-ups \cite{karah2021metanarrative}. These are not compensations but creative capacities embedded in a different cultural grammar of communication. They help articulate a world in which design does not begin with image but with relational attunement. A prototype, here, might be a song pattern, a shared haptic memory route, or a dialogue performed repeatedly over time. This is where the visual culture of prototyping reaches its limit, and blindness helps us imagine beyond it.

\subsection{From Gaze to Stare}
Prototyping often entails watching how \textit{users} behave, predicting their interactions, and interpreting their responses, under a framework that quietly normalizes observation as knowledge. But as blind disability studies scholars have noted, the gaze is never innocent. It is laden with judgment, power, and histories of surveillance. \citeauthor{garland2006ways}'s theorization of the “stare” helps articulate how disabled bodies become sites of excessive scrutiny and meaning-making. The stare, unlike the glance, lingers. It presents people being observed as a spectacle. Within prototyping practices, we note the reproduction of this logic: \textit{user} feedback is often gleaned primarily through visual means, such as screen-capture analysis, emotion journey mapping, or behavior tracking, where the sighted \textit{user} becomes legible precisely by conforming to predictable patterns of interface engagement. In contrast, those who engage systems in non-sighted ways are marked as unpredictable, difficult to observe, and harder to extract design insights from. 

Karah's analysis of the baroque stare versus the blank state offers more than a typology of looking; it reveals the inherent violence embedded in “neutral” observation \cite{epwStaresBlind}. The author describes the baroque stare as “unapologetic and flagrant visual curiosity” that “respects no social, logical, and rational boundaries,” exposing how design research often operates through this same predatory gaze, consuming respondents as visual data while claiming objectivity. But what strikes as particularly significant is how the framework reveals the impossibility of neutral observation in prototyping contexts. The question isn’t whether we’re going to look at \textit{users}, but how our looking is already structured by power relations. When prototyping methodologies insist on “\textit{user} observation” or “behavioral analysis,” they’re often implementing what amounts to a baroque stare, an overindulgent visual consumption that treats human experience as data to be extracted rather than knowledge to be engaged with. 

What emerges from a critical engagement with this literature is the recognition that visual epistemologies are not simply biased toward sight; they are structured through relations of domination that privilege physiological notions of sight as legible and laden with  authority. Michalko’s reflection on entering the classroom with his guide dog reveals more than an instance of discrimination. He describes how blindness precedes him, as an “echo,” a presence shaped by absence, signaling how academic spaces treat blindness not as a form of knowledge but as its negation \cite{michalko2001blindness}. This absence isn’t incidental; it is structurally necessary to sustain the illusion of objective, disembodied knowledge. In this context, prototyping practices in academia inherit similar exclusions. Framed as “human-centered” or “empathetic,” they often enact what Kleege critiques as the “dead in the eye” gaze, a way of looking that turns the act of observing into an ableist performance \cite{kleege1995here}. These practices reproduce ableist norms under the guise of inclusion, making it all the more urgent to rethink who gets to define what counts as knowledge in design. Consider the standard prototyping setup where the designer sits, often literally in a quiet corner, carefully watching a \textit{user} navigate through a screen-based mockup. The \textit{user}, illuminated by screen light and is often filmed by positioned cameras and are later revealed as a part of design success stories. This setup grants the designer epistemic privilege, the power to “see” without being “seen”, to interpret without being questioned. The facilitator guides them through predetermined tasks while the moderator ensures procedural compliance, keeping a check on the time and conversations to be professional. In a way, the act of prototyping requires \textit{users} to be knowledgeable through visual consumption to maintain its authority as a problem-solving methodology.

Such scenes embody the asymmetries of power that cultural theorists like \citeauthor{foucault2020power} describe as the “object of information, never a subject in communication.” Drawing from \citeauthor{butler2009performativity}'s idea of the performative nature of power relations, the \textit{user} is compelled to perform their “user-ness” under surveillance, while the designer performs their authority through invisible observation. Thus, prototyping cannot be disentangled from the politics of the gaze. And in dismantling the gaze, blind scholars offer alternative imaginaries. Prototyping is treated as co-presence rather than control, contributing to the idea that different arrangements might produce different forms of knowledge and different relationships between designers and respondents. Interpretation, then, becomes a shared negotiation moving away from the unilateral act of designer authority. This fundamentally reorients the philosophy of design, in which the prototype is no longer a fixed object to be observed and refined through controlled interaction, but a “social occasion,” as moments of situated meaning-making, keeping the ideas partially open to live in ambiguity and dialogue.    

\section{Reflections and Future Contributions}

In \textit{The Wonderful Wizard of Oz}, transformation does not arrive through the wizard's magic or the good witches' blessings, but through the protagonists' journey of care, relation, and self-discovery. The Tin Woodman, though he believes he lacks a heart, shows deep love and loyalty. The Scarecrow, who perceives himself as brainless, offers steady reasoning and inventive solutions. Their journeys reveal that emotional depth and intelligence are not confined to normative bodily functions but emerge through lived experience, vulnerability, and collective action.

This allegory illuminates the limits of dominant design practices, particularly in Human–Computer Interaction, here human-centered design (HCD) continues to idealize a particular kind of “user.”  While HCD has improved usability by centering \textit{humans}, it has also normalized the idea of a singular, self-evident \textit{user}, flattening diverse embodiments and access needs into a narrow frame. This paper has argued that the logic of \textit{fidelity}, the pursuit of ever-closer approximations to an imagined "real," reproduces exclusion by obscuring whose realities are privileged. What we glimpse, if we listen, is not another version of fidelity, but a refusal to need it. Our future work will continue investigating these access-building practices, not only within academic frames but in the lived, infrastructural, and pedagogical realities of blind learners.

\bibliographystyle{ACM-Reference-Format}
\bibliography{references}

\appendix
\end{document}